# Design of porthole aluminium extrusion dies through mathematical formulation

Juan Llorca-Schenk[a], Irene Sentana-Gadea[a], Miguel Sanchez-Lozano[b]

[a] Department of Graphic Expression, Composition and Projects, University of Alicante, Alicante, Spain
[b] Department of Mechanical Engineering and Energy, Miguel Hernandez University, Elche, Spain

**Abstract**

A mathematical approach to solve the porthole die design problem is achieved by statistical analysis of a large amount of geometric data of successful porthole die designs. Linear and logarithmic regression are used to analyse geometrical data of 596 different ports from 88 first trial dies. Non-significant variables or high correlated variables are discarded according to knowledge of the extrusion process and statistical criteria. This paper focuses on a validation model for a typical case of porthole dies: four cavities and four ports per cavity dies. This mathematical formulation is a way of summarizing in a single expression the experience accumulated in a large number of designs over time. A broad way of research is open to generalise this model or extend it to other types of porthole dies.

## 1. Introduction

Nowadays, the utilisation of aluminium profiles is becoming wider and wider in distinct fields because of its excellent physical properties. Direct extrusion of aluminium is a complicated metal deformation process, involving shape deformation, heat transferring and a complex friction state [1].

Productiveness, effective cost and quality grade of the extruded profiles are the main commercial factors. These three factors are directly linked to the performance of the extrusion die. In addition, there are some other factors such as extrusion press features, billet material quality, auxiliary equipment capabilities, and latter operations such as age hardening, painting, anodizing… Because of its very fine tolerances, special material and high demands on thermo-mechanical fatigue performance, the die probably is the most critical extrusion component [2].

Extrusion of hollow profiles is a quite usual industrial process conducted using so-called porthole dies. The latter consists of two different parts, concretely the mandrel and the die plate. The bearing, which is the shape that forms the profile, is divided over these two elements. For hollow profiles, the die plate gives the shape of the outer contour of the profile. Likewise, the inner contours of the profile are shaped by the external contour of the mandrel. For the purpose of allowing the aluminium to flow from the front end of the mandrel to the bearing zone, so-called portholes are milled in the mandrel. Steel zones of the mandrel between the portholes are named bridges. In this manner, portholes function is to allow the aluminium flow through the die, while bridges function is to fixate the position of the mandrel core. During the extrusion process aluminium flow is split by the bridges and forced through the portholes. After overcoming the bridges, the aluminium flow should weld together again in the so-called welding chamber. This weld occurs in solid state conditions at proper pressure and temperature conditions [3].

The traditional way of extrusion dies design is mostly founded on analogy engineering and similar previous design experiences. Numerical simulation by the finite elements method (FEM) can be used for the aluminium extrusion process but the application of this kind of calculations in the extrusion industry is limited because of the high complexity involved [1].

Empirical design approach is another popular way for extrusion die design. During past years, a lot of design rules and formulas have been introduced and studied. Some authors address specific issues like: Bearing length and layout design [4], construction factors and bearing length design guidelines [5]…

Looking at the scientific literature, few works were presented in the last years regarding specifically porthole die design fundamentals. Some papers present basic guidelines about different porthole die design points but they are only general guidelines.

Some of the most important papers are from biannual ICEB (International Conference on Extrusion and Benchmark). Event where experts in the field of aluminium extrusion join together in order to verify the increase of accuracy of FEM simulations for process optimization and also to share knowledge advancements on the matter. For these reasons a different die is designed at every benchmark: In 2007 the focus was on pocket design [6], in 2009 it was on the influence of tongue deflection in U shape profiles [7], in 2011 different strategies for port-hole balancing were used for hollow profiles extrusion [8], in 2013 the experimental investigations were aimed at predicting the effects of mandrel deflection [9] while in 2015 the effect of bearing shape (straight, choked or relieved) and length were investigated [10].

The use of finite elements simulation for extrusion dies design is an encouraging possibility for extruders and for tool makers. Commercial software packages provide user friendly interfaces and offer a wide variety of results: differences in extruded profile velocity, tool deflection, tool stress and profile temperature. The feasibility of die design corrections without press trials saves much money and time. One of the drawbacks using FEM simulation is the necessity of simulation experts for the correct preparation and later analysis. Depending on the difficulty of the die design, preparation and analysis can take hours or days without counting the time spent for the numerical calculation. Like so, this decreases the attractiveness of FEM simulations for extrusion processes on the account of raised design times, extra software costs and extra staff costs [11].

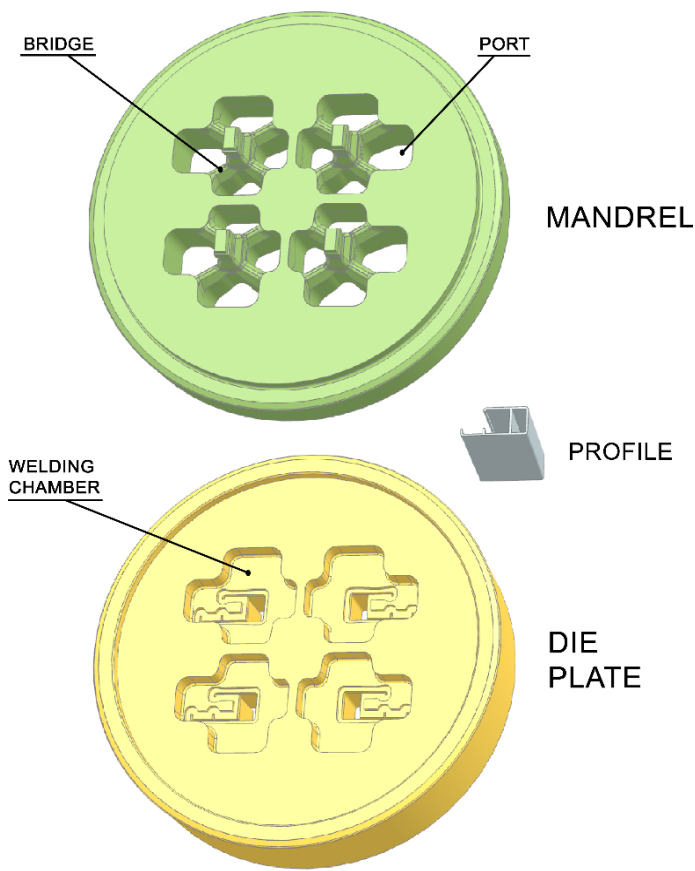

**Fig. 1.** Example of mandrel and die plate of a disassembled porthole die

Considering the difficulties and costs of FEM simulation, this paper tries to offer a mathematical tool to the designer for successfully dimensioning ports of porthole extrusion dies (see fig. 1). Wrong geometrical properties of ports (area, position…) are the most important source of serious problems during extrusion in porthole dies, which can produce a large deflection of the extruded profile or big lateral displacements of the mandrel and modifications of profile thicknesses.

As has already been mentioned, there are many papers offering design guidelines, recommendations for die design and some specific formulas to define some particular detail in porthole dies. But none of them provides a structured and precise formulation to facilitate port design and dimensioning for porthole extrusion dies. Apparently, in the industrial world some large companies and renowned die makers have some sort of these tools to assist designer in ports geometry definition but all this knowledge belongs to its know-how and there is not any publication on this matter.

It could be generalized that, to obtain an optimal porthole die design it is necessary to ensure: the mechanical resistance of all die pieces and a uniform exit velocity in the aluminium profile during the extrusion process. Resistance calculations allow guaranteeing optimal mechanical properties. The principal design variables to achieve a uniform exit velocity in porthole dies are: balanced ports geometry definition to allow a uniform aluminium flow in the welding chamber, optimal welding chamber definition [12] and optimised bearing definition depending on profile thickness and its position in the die.

Nevertheless, to define balanced ports is not so easy because velocity distribution inside aluminium billet is not uniform. Due to the friction between billet and container wall, aluminium pressure [13] and velocity distribution at the front end of the extrusion die are concentric (its maximal velocity is in the centre and the minimal velocity is in the external zone).

The new tool presented in this paper intends to be used as definitive designing assistance to define balanced ports geometries in simple or medium difficulty porthole dies. Also, it could be used as starting point for designing high difficult porthole dies to reduce number of iterations of FEM simulation/modifications to achieve an optimal solution.

## 2. Materials and Methods

### 2.1. Overview of the method

The general objective of this research is to specify a tool to help die designers to obtain balanced ports geometries for porthole dies.

But, when could we consider balanced all the ports of a porthole die?

The most extended design criterion is achieving to equal aluminium concentric velocity differences from the front end of the die to the end of the trajectory inside the port. With this criterion, after an optimal balanced design of ports, bearings definition is quite simple because it mainly depends on the profile thickness.

The problem of designing porthole dies ports appears mainly in dies of several cavities because they have some ports close to the centre of the die and some others far from the centre of the die. Given the concentric distribution of pressure inside the container [13], the ports must be dimensioned so that their area/perimeter ratio depends on their relative position with respect to the centre of the die.

Apparently, it seems extremely difficult to obtain an application that automatically generates balanced port geometries from scratch because ports geometry must be adjusted to profile geometry and depends on other several factors. An alternative possibility could be to define a tool that enables validating new ports geometries of porthole dies, ensuring a balanced ports design.

For experienced designers it seems intuitively plausible the existence of a mathematical balancing function which links the different geometrical variables of the ports (area, perimeter and distance to die centre) with geometrical variables of the profile zone influenced by each port.

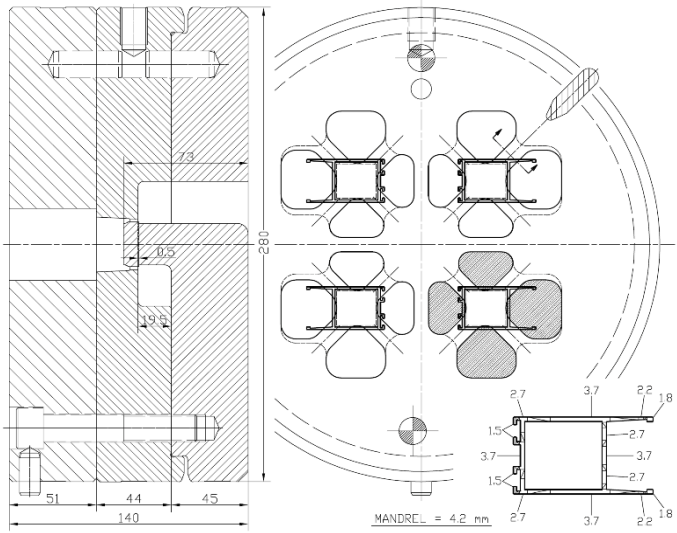

**Fig. 2.** Example of a typical four cavities porthole die design with four independent ports for each cavity (hatched port geometries in one of the cavities).

Given the wide variety of die designs, after doing some tests it has been determined that the best way to achieve optimal results is to group die designs by different typologies. As the objective is to obtain a balanced ports design, chosen typologies depend on the die number of cavities and the number of ports for each cavity.

This research is centred in four cavities porthole die designs with four independent ports for each cavity (figure 2). This is probably the most common four cavities type of porthole die design and it is a widespread die typology for medium sized profiles for any application.

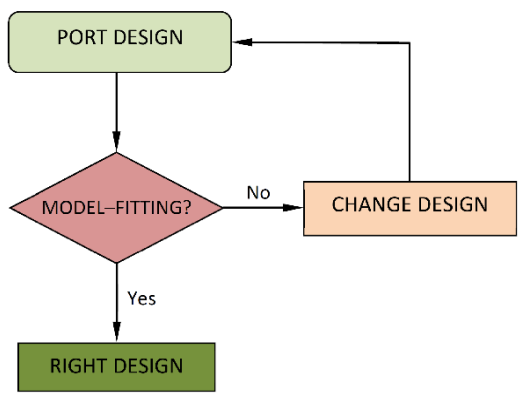

**Fig. 3.** Diagram of the design methodology.

Now it is worth considering, could a univocal mathematical function between the geometrical variables of balanced ports help designers to obtain an optimal die design?

The designer develops an initial ports design and check if these designed ports conform to the mathematical formulation. If designed ports don't conform to the model with a previous defined tolerance, ports design must be modified in the correct sense to approach a balanced situation. These modifications and conformations have to be repeated until achieving a port configuration that conform to the mathematical formulation with some tolerance (figure 3).

What is the methodology used to obtain this univocal function that relates the geometrical variables of a balanced port design for a porthole die?

The methodology is based on performing a statistical regression analysis in order to obtain a function that relates the value of some variables for the balanced ports. The analysis has focused fundamentally on the search for a regression that maximizes the variance of the data covered by the model and does not present important signs of correlation between the independent variables.

From 88 tried and tested four cavities and four ports per cavity porthole die designs, a huge number of geometrical variables of ports geometry and profiles geometry are analysed. They are all dies used on 8-inch 22000MN or 7-inch 16000MN presses. These two types of presses have a very similar commitment because both have practically the same maximum pressure. Many extruders simultaneously have a press of each of these types and typically use the dies from the 7" press on the 8" press with equivalent port balancing results.

The proven efficacy of these 88 dies is determined because they are all fist trial dies produced by HYDRO ALUMINIUM EXTRUSION PORTUGAL HAEP, S.A. After first trial, die behaviour was optimal and profile samples were approved in accordance to manufacturing tolerances fixed by EN-12020-2 (according to feedback information after trials)

*2.2. Analysis variables*

The first step to define this new design assistance tool is determining what variables are fundamental for geometrical ports definition in a porthole die. Initially, this high number of geometrical properties is collected for each port:

1. Port area.
2. Port perimeter.
3. Distance from port centre to die centre (distance from areas' centre of port geometry to die centre).
4. Area of aluminium profile zone affected by related port (see Note 1 and figure 4).
5. Perimeter of aluminium profile zone affected by related port (see Note 1 and figure 4).
6. Total area of all the die ports.
7. Total perimeter of all the die ports.
8. Distance from the areas' centre of port to areas' centre of the aluminium profile zone affected by related port (see Note 1 and figure 4).
9. Depth of the port or bridge height.

10. Container diameter of the press.
11. Maximum pressure of the press.

**Note 1:** A certain portion of the extruded profile is shaped by the aluminium getting through each port. This profile portion is delimited by welding lines, determined in die design by bridge geometry [14].

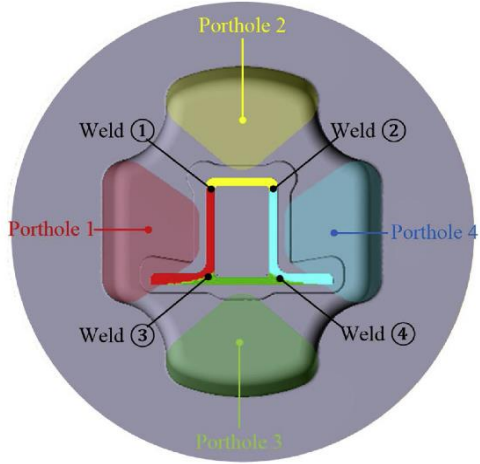

**Fig. 4.** Material distribution in ports and its corresponding welds and distribution in profile geometry. Source: [14]

A brief explanation of each of these variables is listed below with the support of some figures. In figure 5 it is possible to see the three most important geometrical variables of a porthole die port from the point of view of the design for an optimal extrusion process.
- *Port area* predetermines the aluminium quantity getting to that zone of the die
- *Port perimeter* conditions is restraining capability to aluminium flow through it

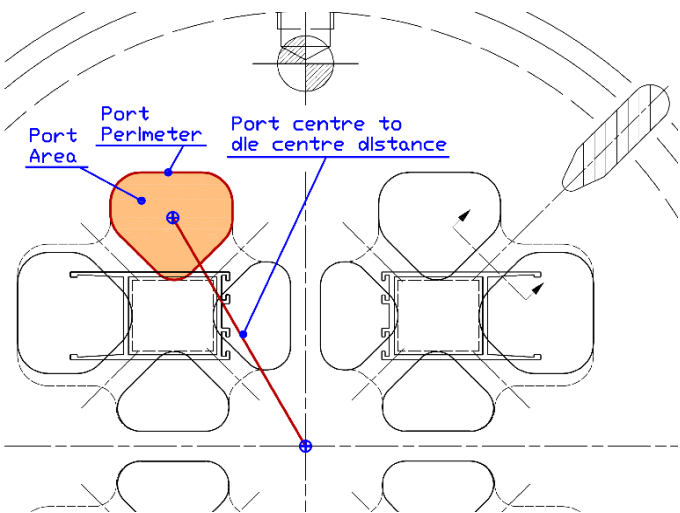

**Fig. 5.** Most important geometrical variables of a porthole die port from the point of view of the design.

- *Distance from port centre to die centre* predetermines the aluminium velocity in the front end zone of the port. In fact, the exact location is not important but only its distance to the centre. Thus, symmetrical ports with respect to die centre (same geometry and symmetrical position) show the same behaviour during extrusion process.

To try to take into account the influence of the profile portion affected by each port, the following variables have been taken (fig. 6):
- *Area of aluminium profile zone affected by related port* predetermines the amount of aluminium that should exit through that zone of the die. Therefore, it could also condition the amount of aluminium that must enter inside the port.
- *Perimeter of aluminium profile zone affected by related port* conditions the restraining capability of the profile to the free flow of aluminium in that zone of the die. Therefore, it could also condition the necessary aluminium speed that must enter inside the port to be balanced with other ports of the die.
- *Distance from areas' centre of port to areas' centre of aluminium profile zone affected by related port* predetermines how the profile is directly exposed to the aluminium flow inside the port.

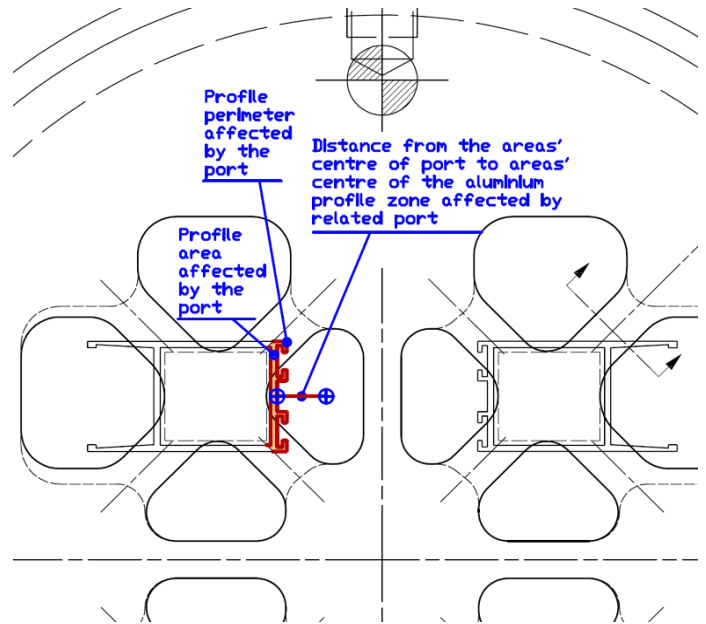

**Fig. 6.** Other geometrical variables of a porthole die used in this analysis.

Also, two geometric variables that correspond to the complete set of ports are included because, experience indicates that the same die can be balanced with larger or smaller ports. Therefore, these general variables serve to integrate in the regression if the design uses larger or smaller ports. These variables are:
- *Total area of all ports of the die* predetermines the general amount of aluminium to flow inside the die.
- *Total perimeter of all ports of the* die conditions the general restraining capability to aluminium flow through the die.

Finally, three additional variables characteristics of the die and the extrusion press have been included:

- *Depth of the port* determines how much the aluminium is slowed as it flows through the port.
- *Container diameter of the press* determines, together with the maximum pressure of the press, the distribution of pressures in the die.
- *Maximum pressure of the press* determines, together with the container diameter of the press, the distribution of pressures in the die.

To facilitate the collection of data, a C# application for AutoCad has been developed to capture the data from 2D die designs in DWG format. This application facilitates the task through the automatic selection of the layers and the orderly writing of the data of the different geometries to be taken into account.

All data are collected in millimetres and square millimetres.

*2.3. Analysis definition*

The area of the port has been chosen to be defined as the dependent variable, because it is the variable that is usually used as a characteristic identifier of each port (the perimeter and the distance to the centre also define the port geometrically but normally the changes and adjustments of the ports are made by changing its area).

The software IBM SPSS 26.0 was used for all linear regression analyses performed. The Stepwise method was used, which automatically introduces and extracts variables from the model according to the critical statistical significance levels of Student's t distribution.

The t-tests and their critical levels serve to contrast the null hypothesis that a regression coefficient is worth zero in the population. Small critical levels (less than 0.05) indicate that this null hypothesis should be rejected and therefore indicate that the variable has a significant weight in the regression equation.

The Stepwise method leaves out of the model those variables that show a critical level greater than 0.10 (rejection probability). Thus, the method uses several iterations to finally obtain a regression model that maximizes the fit of the model and minimizes the number of significant variables used.

The Adjusted Coefficient of Determination (denoted *adjusted* $R^2$ or $R^2_{adj}$) defines the proportion of the variance in the dependent variable that is predictable from the independent variables.

Standard Error is a variability measure of the dependent variable part that is not explained by the regression equation.

In order to try to verify the independence of the variables, the Partial Correlation Coefficient expresses the degree of relationship between two variables after eliminating, from both, the effect due to third variables. Therefore, they express the degree of relationship existing between each independent variable and the dependent variable after eliminating from both the effect due to the rest of independent variables included in the regression equation.

It is also possible to verify independence through the Semi-partial Correlation Coefficient, which expresses the degree of relationship between two variables after eliminating from one of them the effect due to third variables. Therefore, the semi-partial correlation coefficients express the degree of relationship between the dependent variable and the part of each independent variable that is not explained by the rest of the independent variables included in the regression equation.

And in order to quantify the importance of each of the independent variables in the regression, the Typified Regression Coefficients (Beta coefficients) will be used.

The next step could be to try to reduce the number of fundamental variables to be used in the definition of the objective function using the following criteria:
- The knowledge of the extrusion process
- The typology of the geometries used in the design of the ports
- The preliminary results of the first statistical analyses carried out with the total set of all the data.

**3. Results**

The first conclusion after first statistical analyses is that port Perimeter independent variable has a very high partial correlation with the dependent variable port Area. This means that the degree of relationship between Perimeter and Area after eliminating, from both, the effect due to the rest of the independent variables included in the regression equation is very high.

Physically it is possible to explain the great correlation between the Perimeter and the Area of the ports. In fact, you only have to look at the examples of designs that have been shown so far and it is possible to see that the ports have different shapes and sizes but all of them have in common a typology of similar shape.

In other words, there is a quite direct relationship between the Area and the Perimeter of the ports because their shape is always similar.

Therefore, it is most reasonable to discard the Perimeter variable as an independent variable because the relationship existing with the Area (dependent variable) is not due to extrusion phenomena but to a simple geometric linkage. For the same reason, Total Perimeter is removed from the group of dependent variables, maintaining Total Area.

The other conclusion from the first statistical analysis is that certain variables are far from being statistically significant. These are:
- Depth of the port. Due to the type of dies chosen (four cavities and four ports per cavity), this variable is not shown to be significant because the depth values are very similar. All of them are in the range of 40-50mm
- Container diameter of the press and maximal press pressure. For this analysis, dies have been chosen from two types of presses with very similar behaviour. Consequently, these variables do not show any statistical significance.

Finally, two different alternative models have been obtained from linear regressions. A first model based on a linear regression from the commented variables and a second model based on logarithms of the commented variables. In this second case, the logarithms are used to obtain a non-linear model by means of a linear regression.

*Linear model:*

From this regression,

$$Area = -25.048 + 5.072\, Dist + 0.012\, Area_{Total} + 0.593\, Area_{Prof} + 10.358\, Dist_{Port\text{-}Prof} + 1.211\, Perim_{Prof} \quad (1)$$

It is possible to obtain this verification formula:

$$0 = -25.048 - Area + 5.072\, Dist + 0.012\, Area_{Total} + 0.593\, Area_{Prof} + 10.358\, Dist_{Port\text{-}Prof} + 1.211\, Perim_{Prof} \quad (2)$$

*Non-linear model:*

From this regression,

$$Ln(Area) = 0.956 + 0.479\, Ln(Dist) + 0.304\, Ln(Area_{Total}) + 0.111\, Ln(Perim_{Prof}) + 0.120\, Ln(Dist_{Port\_Prof}) \quad (3)$$

It is possible to obtain this verification formula:

$$1 = 2.6013 \cdot (Dist)^{0.479} \cdot (Area_{Total})^{0.304} \cdot (Perim_{Prof})^{0.111} \cdot (Dist_{Port\_Prof})^{0.120} / Area \quad (4)$$

Where each variable is:

- *Area* - Port area (in mm)
- *Dist* - Distance from port centre to die centre (distance from areas' centre of port geometry to die centre) (in mm)
- $Area_{Prof}$ - Area of aluminium profile zone affected by related port (in mm$^2$)
- $Perim_{Prof}$ - Perimeter of aluminium profile zone affected by related port (in mm)
- $Area_{Total}$ - Total area of all ports of the die (in mm$^2$)
- $Dist_{Port\_Prof}$ - Distance from port areas' centre to areas' centre of aluminium profile zone affected by related port (in mm)

In the dataset associated with this article [15] it is possible to see the summary tables of the regressions obtained by means of IBM SPSS.

These two regressions (1)(3) have been selected among all those tested because they present the following benefits:
- They maximize the degree of fit between the model and the dependent variable.
- They minimize the value of the estimate typical error. That is, they minimize the typical error of the regression residues.
- They do not present important correlation problems between independent variables.

Both models show a very similar variable adjustment, the Adjusted Coefficient of Determination (denoted *adjusted* $R^2$) is practically the same for them: 0.778 in the linear model and 0.780 in the non-linear model.

On the other hand, the linear model has the advantage that it is much easier to define a working tolerance for the intended mode of use.

For these reasons, it is believed that the most interesting and practical solution is to use this linear model verification formula (2).

It is essential to set some working tolerance because almost always after checking the model it is impossible for all the ports to get exactly zero as a result of the linear model verification. An acceptance tolerance must be defined to validate the checked value obtained.

Observing the results obtained for the regression corresponding to the verification linear model, the standard error of the Area estimation is 70.77mm$^2$. Therefore, it could be said that the model fits the values of the area obtained ± 70mm$^2$.

Taking as a reference half of the standard error, to use the model, it would be possible to check for all ports if the following mathematical expression (5) is met, in which a maximum tolerance of ±35mm$^2$ is allowed.

*Definitive model:*

$$35 > -25.048 - Area + 5.072\, Dist + 0.012\, Area_{Total} + 0.593\, Area_{Prof} + 10.358\, Dist_{Port\text{-}Prof} + 1.211\, Perim_{Prof} > -35 \quad (5)$$

If the values obtained for the verification formula are between -35 and +35 for each port, the design being analysed would be considered valid.

Looking at the two models, it is possible to reflect on the physical validity of the results obtained. Although the two models are clearly different, they both have in common the type of contributions and the weight of each of the variables within the model:

- Observing the Beta Coefficients corresponding to each one of these models, it is noted that in both cases the independent variable with the greatest weight is the Distance and is followed by the Total Area.
- The rest of the variables have similar weights in both cases.
- The only notable difference is that the variable with the lowest weight of the model, Profile Area, has not been shown to be significant in the non-linear model. Therefore, this variable does not participate in this model (its weight is very low in linear model)

Next, it seems logical to ask whether the type of contribution and the meaning of each contribution is what would be physically expected in an extrusion process. To clarify this issue, it is detailed below what the direction of the contribution should be for each of the variables according to the logic of extrusion:

- The relation *Port Area / Distance from port centre to die centre,* is strong with a positive sense. In order to achieve a balanced flow in the die, the sense is positive. It is necessary to increase the port area if port distance to the centre is greater. The concentric distribution of pressures in the die [13] causes this need. The relationship is strong because the decrease in pressure with distance is rapid and must be balanced with the increase in area.

- The relation *Port Area / Total Area of ports in the die* has a positive sense. Evidently, there is some correlation between these two variables because one of them participates in the other one. But the experience of extrusion indicates that the same die can be balanced with larger or smaller set of ports. This is why, it has been initially considered convenient to include two global variables in the analysis: *Total Area* and *Total Perimeter* of the ports.
- The relation *Port Area / Profile Perimeter* has a positive sense. The greater the perimeter of the profile portion fed by a port, the greater the area of the port must be to achieve a balanced flow. A greater perimeter of the profile represents a greater hindrance to the aluminium flow. In order to overcome this hindrance, a higher pressure of the aluminium is necessary, which is achieved by means of a larger port area.
- The relation *Port Area / Profile Area* has a positive sense. The larger the area of the profile portion fed by a port, the larger the port area must be to achieve a balanced flow. A larger profile area represents a greater need for aluminium volume. But experience shows that this dependence is only met for large differences in the profile area. Therefore, the influence of the profile area is the smallest and in the non-linear case has not even been shown to be significant.
- The relation *Port area / Distance from the centre of profile areas to the centre of port areas* also has a positive sense. The *Distance from the centre of profile areas to the centre of port areas* is a way of representing whether the profile is more or less directly exposed to the flow of aluminium entering the port. The smaller this distance, the more directly the aluminium will flow out of the profile. Therefore, the greater the distance, the more difficulties the aluminium will have in its flow and the greater its area should be to achieve a balanced flow with respect to the rest of the ports.

Therefore, it is confirmed that the relationships between the different independent variables and the dependent variable are those expected in the models according to the logical reasoning of the aluminium behaviour during extrusion.

This verification ensures that the expressions obtained can be assumed to have been validated from a theoretical point of view. The next step should be a practical example of application to show off the goodness of these expressions to facilitate the correct dimensioning of a new die ports.

## 4. Application example

### 4.1. Verification formula application

The way to apply the solution model to the application example follows the scheme shown in figure 3. The different steps of the process are detailed below:

1. Firstly, from the profile geometry required to manufacture a die, a design of the set of ports is created according to the design criteria imposed by the profile geometry and the designer's experience. (In this step of the process the mathematical model does not intervene at all)
2. For each of the ports designed, the value of the verification formula of the linear model is evaluated.
3. If the values obtained for the model in all the ports are inside the range [-35, +35], the design will be validated.
4. If model value for any port is outside the range [-35, +35], the design of that out of range port must be modified.
5. After this, the verification formula for all ports is evaluated again (if any port changes, the value of the model for the rest of ports changes because the *Total Area* variable participates in the model). If any port continues to present model values outside the verification range, step 4 must be repeated until all ports fit the model.

If it is necessary to make modifications to the ports to fit the model, there are two variables on which it is possible to act: The *Port Area* and the *Distance from port centre to die centre*. In other words, it is possible to change the size of the ports and/or the position of the ports. But, in principle, a change in the area of the port is usually simpler and quicker because the position of the profile in the die is usually more or less fixed by other design conditions. Therefore, the most interesting method to modify the parameters of a port is to modify its area by changing the area furthest away from the profile. (In the example it is possible to see that it is the fastest way from the point of view of the tools available in CAD applications for the type of geometries that ports usually have).

Figure 7 shows the profile used in the application example and the desired extrusion position for it. This is a simple profile for which a design of four cavities and four ports per cavity is desired for an 8" 2200MN press.

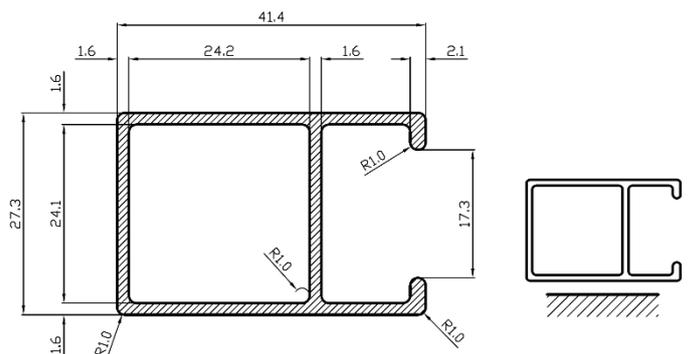

**Fig. 7.** Profile used in the application example

From this data, the designer would begin to make the design of the die. Among the different steps involved in the design of this die, we will focus on the design of the ports of the mandrels, for which the mathematical model has been developed.

Initially, the designer would position the profile cavities with respect to the centre of the die, following the usual extrusion criteria (symmetry and customer indications) and with the conditioning of the bolster or insert support.

Next, the designer would make a first attempt to design the ports taking into account the position set for the cavities and the need to leave between ports a minimal space of between 10-13mm (usual bridge width range for dies of four cavities, although a resistance calculation of the bridges must always be made)

Under these conditions, the designer decides to perform the initial design (fig. 8). This is a symmetrical design with bridges of 12 mm in which, at first sight of a person with experience in porthole die design, it could be thought that the ports furthest from the centre seem slightly oversized.

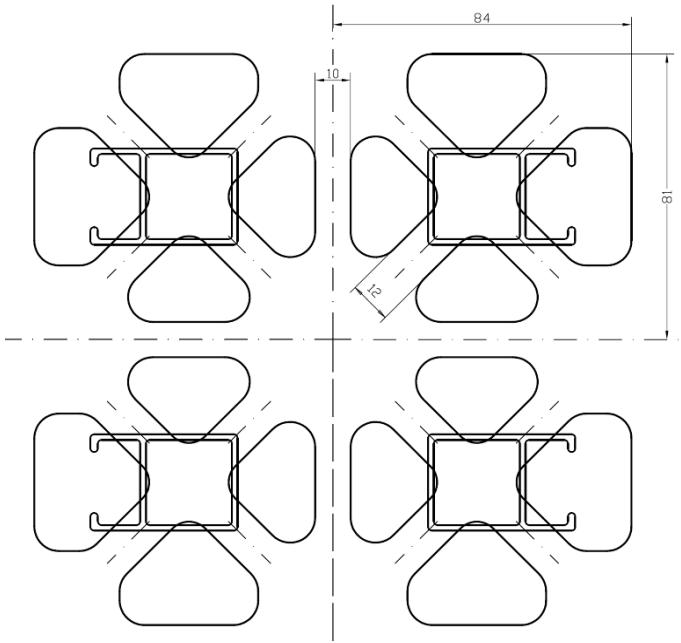

**Fig. 8.** Initial design for example die

For these initial ports, the value of each of the variables that participate in the model is shown in figure 9.

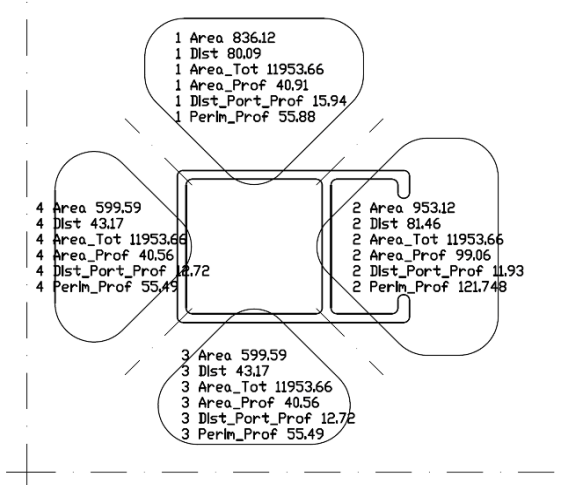

**Fig. 9.** Initial value of each of the variables

Figure 10 shows the verification formula values obtained for each of the ports.

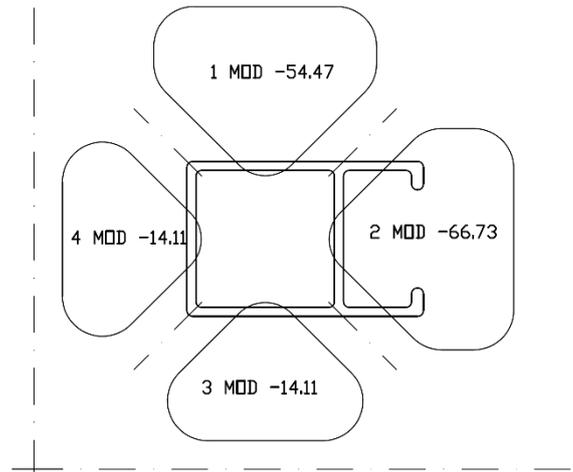

**Fig. 10.** Initial verification formula values for each port

According to the results obtained, the internal ports (numbered 3 and 4) meet the model with the defined tolerance range [-35, +35]. But external ports (numbered 1 and 2) are clearly out of the range

With the obtained values it is possible to know in which direction the port must be modified to achieve a greater approximation to the valid range of the model:
- Ports whose model value is less than -35 must be modified in a way that their area is reduced or their distance to the centre is increased (it has already been commented that it is much easier and more practical to modify the area than to try to move the port centre).
- Ports whose model value is greater than +35 must be modified in a way that their area is increased or their distance to the centre is reduced.

Therefore, the area size of ports numbered 1 and 2 should be reduced in order to try to get the model to the desired values.

One difficulty with this methodology is that the model does not provide clear information about how much the area of the ports should be modified.

Since other variables are linked to changes in the port area, it is impossible to quantify exactly how much the port area needs to be modified, only one approximation can be estimated.

The order of magnitude in which the area must be modified would be given approximately by the distance between the value obtained for the model and the nearest limit of the valid range for the model [-35, +35]. For example, in the previous case, the value of the model for port 1 is -54.47; therefore, the minimal decrease of its area should be approximately: (-54.47) - (-35) = -19.47

Or even the distance to the centre of the range, if the zero value for the verification formula is to be met:

(-54.47) - 0 = -54.47

For each of the ports, the order of magnitude of the modification to be made to the area should be at least:
- Port 1: -54.47- 0 = -54.47, a 6.5% reduction.
- Port 2: -66.73 - 0 = -66.73, a 7% reduction.
- Port 3: no change is necessary
- Port 4: no change is necessary

With these premises, the designer should modify the size of the ports trying to fit as closely as possible to the model. If this task is performed for the example port design, a new design is reached (fig. 11).

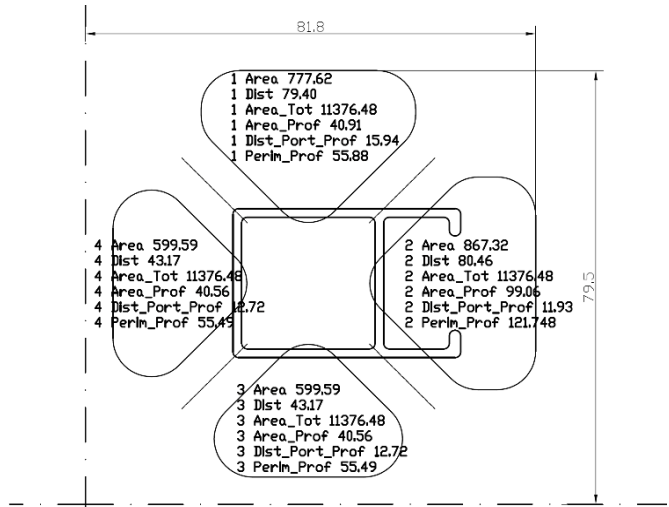

**Fig. 11.** New variables values after die design modification

Figure 12 shows the verification formula values obtained for each of the balanced ports.

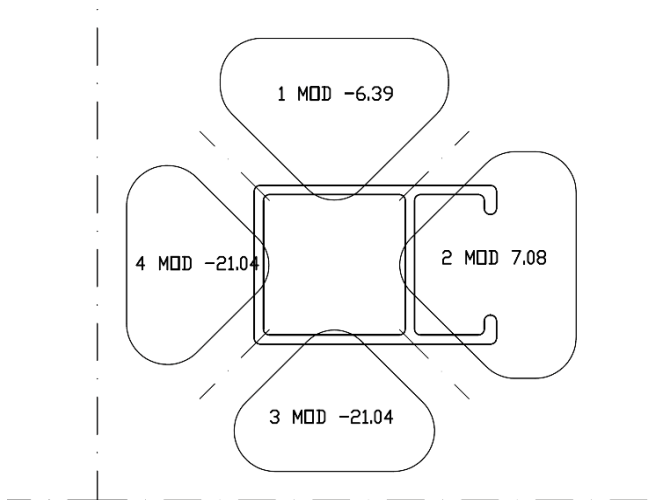

**Fig. 12.** New verification formula values after die design modification

In this way, all ports conform to the conditions imposed by the model because the value obtained is within the range [-35, +35]. Therefore, this port design could be considered valid.

Once the ports design is finished, the designer should continue with the design of the die and complete the rest of the necessary elements to finish defining it. Once the ports are balanced, the definition of the bearings only depends on the thickness of the profile and the position of the profile in relation to the bridges. Finally, the complete design of the die is shown in figure 13.

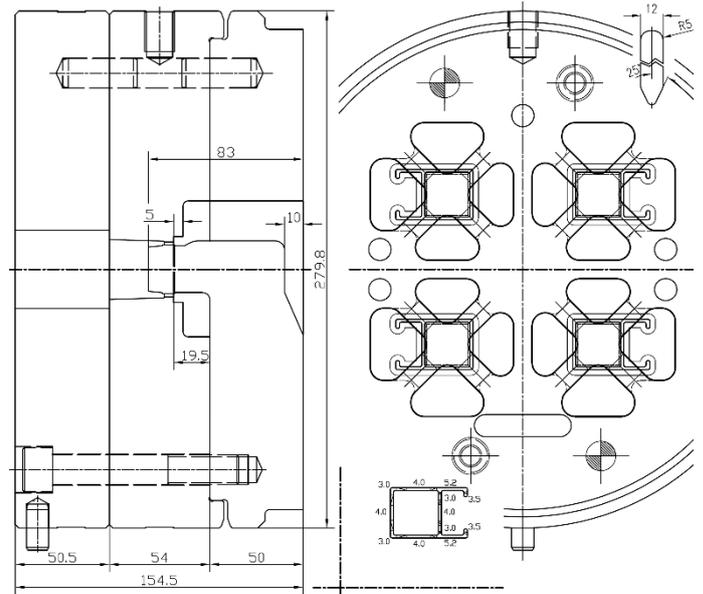

**Fig. 13.** Complete example die design

### 4.2. Verification with FEM simulation

To verify the model predictions, they are compared with numerical simulation predictions employing the Qform Extrusion software by QuantorForm Ltd. It is a special-purpose program for the extrusion simulation based on the Lagrange–Euler approach. This software provides material flow analysis coupled with the mechanical problem in the tooling set, considering shape changes produced by the die deformation on the material flow through the die, which helps to ensure high accuracy of the numerical results [16].

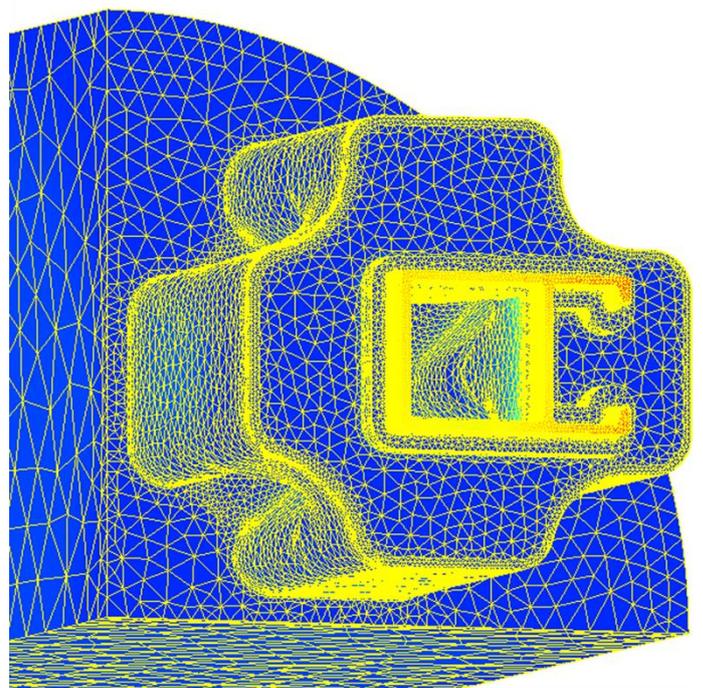

**Fig. 14.** FE workpiece flow domain of initial die design

These figures show the Finite Element Model (FEM) of the aluminium workpiece and die-set. Due to the symmetry of the chosen geometry, a quarter model is created. The workpiece flow domain (Fig. 14) is the volume of aluminium that fills the container and the inner space of the die-set. The die deflection influence on the material flow needs to be considered in the study; for this reason, elements are also assigned in the die-set (Fig. 15).

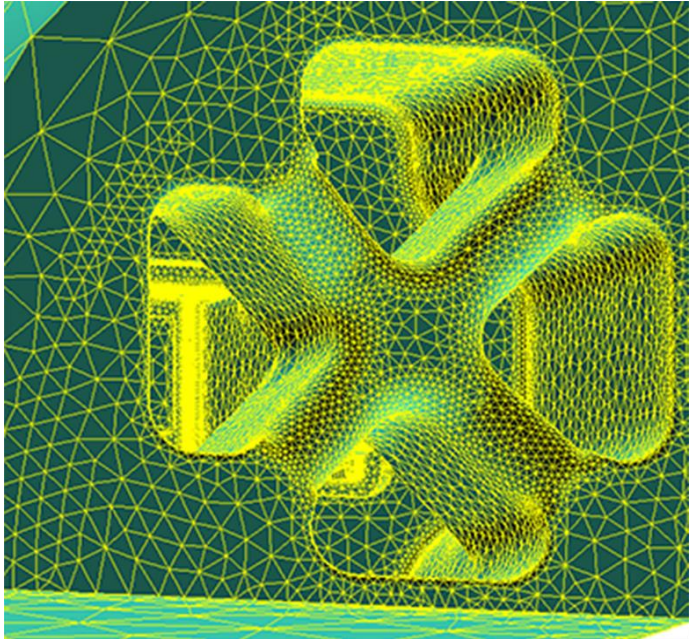

**Fig. 15.** FE die-set model of initial die design

All meshes are built using tetrahedral elements, and the number of volumetric elements in the whole model is 1603305: 985062 elements in the workpiece and 618243 elements in the die-set. The number of nodes in the whole model is 319570: 199623 nodes in the workpiece and 119947 in the die-set.

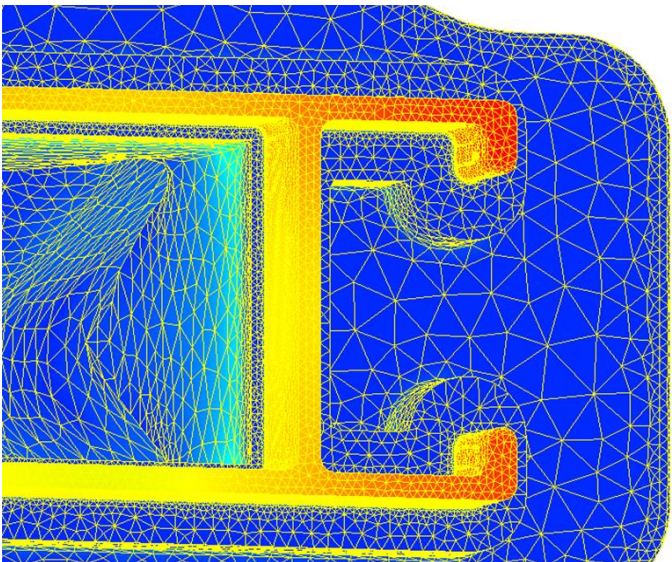

**Fig. 16.** FE model detail of the die bearing region

To control the mesh quality and obtain accurate calculation results, a high-adaptive mesh is applied, and different element sizes in different regions of the whole model are assigned depending on the extent of the local deformation during the extrusion process. Figure 16 shows the region of the die bearing where very fine elements are assigned due to the occurrence of the biggest deformation.

The billet and die materials are EN AW-6063-O aluminium alloy and AISI H-13 steel, respectively. All material properties are temperature dependent.

The die material is considered as an elastic-plastic continuum subjected to small deformations. Its properties are based on different types of functions. Table 1 shows the tabulated values used by Qform Extrusion for the generation of its mechanical properties functions.

**Table 1**
AISI H-13 steel mechanical properties over temperature

| Temperature [ºC] | Young Module [MPa] |
|---|---|
| 20 | 210000 |
| 300 | 187000 |
| 600 | 160000 |
| Temperature [ºC] | Yield Stress [MPa] |
| 20 | 1500 |
| 300 | 1300 |
| 500 | 1100 |
| 570 | 1020 |
| Temperature [ºC] | Density [kg/m3] |
| 20 | 7716 |
| 100 | 7692 |
| 200 | 7660 |
| 800 | 7459 |

Poisson's ratio can be considered constant in relation to temperature. Its value for this steel is 0.3

Table 2 shows the tabulated values used by Qform Extrusion for the generation of AISI H-13 steel thermal properties functions.

**Table 2**
AISI H-13 steel thermal properties over temperature

| Temperature [ºC] | Thermal Conductivity [W/(m·K)] |
|---|---|
| 20 | 22 |
| 300 | 29 |
| 600 | 31 |
| 900 | 32 |
| Temperature [ºC] | Specific Heat [J/(kg·K)] |
| 20 | 375 |
| 200 | 551 |
| 500 | 630 |
| 700 | 975 |
| 800 | 793 |

For EN AW-6063-O aluminium alloy, properties whose variation is based on a linear function are: Density, thermal conductivity, thermal expansion, specific heat, Poisson's ratio and Young module. The tabulated values used by Qform Extrusion for the generation of property functions are shown in table 3.

The heat exchange between aluminium and steel in the contact areas is 30000 W/(m$^2$·K).

**Table 3**
EN AW-6063-O aluminium alloy properties over temperature

| Temperature [ºC] | Density [kg/m3] | Young Module [MPa] |
|---|---|---|
| 20 | 2699 | 70600 |
| 500 | 2586 | 46000 |
| Temperature [ºC] | Thermal Conductivity [W/(m·K)] | Specific Heat [J/(kg·K)] |
| 20 | 205 | 904 |
| 500 | 247 | 1026 |
| Temperature [ºC] | Thermal Expansion [1/ºC] | Poisson Ratio |
| 20 | 2.26·E-5 | 0.33 |
| 500 | 2.78·e-5 | 0.36 |

International Conference on Extrusion and Benchmark (ICEB) recommends the use of the Hansel-Spittel (H-S) model for flow stress modelling in extrusion simulation (6). H-S model allows the flow stress representation by considering also the dependence with strain. The H-S function is obtained from the regression of the experimental data of hot torsion tests [10].

$$\sigma = A \cdot e^{m_1 T} \cdot T^{m_9} \cdot \varepsilon^{m_2} \cdot e^{m_4/\varepsilon} \cdot (1+\varepsilon)^{m_5 T} \cdot e^{m_7 \varepsilon} \cdot \varepsilon^{m_3} \cdot \varepsilon^{m_8 T} \quad (6)$$

Test values used by Qform Extrusion are graphically reported in Fig. 17 at different temperatures and strain rates.

Regression of flow stress data can be performed with all 9 regression coefficients (A and $m_1$ to $m_9$) or with only a part of them. Regression with less than 6 coefficients is not suggested by ICEB due to its low correlation $R^2$-index. Qform Extrusion regression coefficients for EN AW-6063-O flow stress (6 coefficients) are shown in table 4.

**Table 4**
Hansel-Spittel coefficients for EN AW-6063-O aluminium alloy in ºC

| Coefficient | Value | Coefficient | Value |
|---|---|---|---|
| A [MPa] | 265 | $m_1$ | -0.00458 |
| $m_2$ | -0.12712 | $m_3$ | 0.12 |
| $m_4$ | -0.0161 | $m_5$ | 0.00026 |
| $m_7$ | 0 | $m_8$ | 0 |
| $m_9$ | 0 | | |

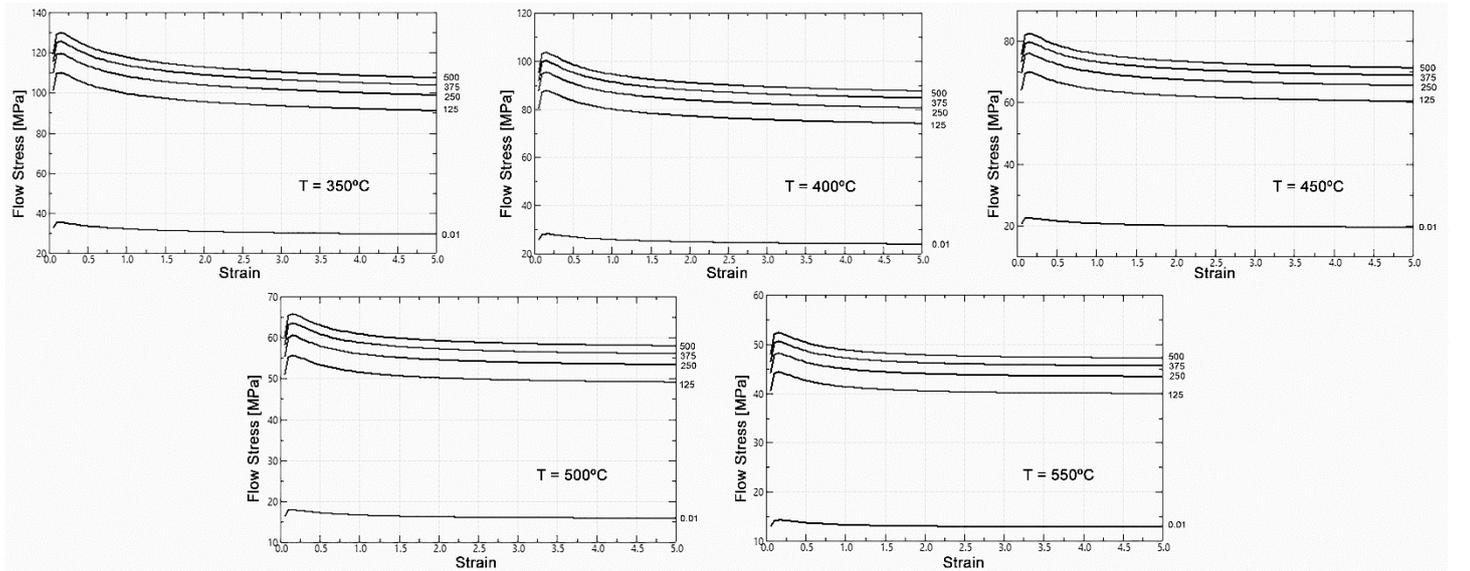

**Fig. 17.** EN AW 6063-O alloy hot torsion test data at 350-400-450-500-550°C temperatures and 0.01; 125; 250; 375; 500 [1/s] strain rates.

Friction of aluminium on steel at extrusion temperatures is usually computed as complete sticking, except in bearing zones where some sliding may occur. It is suggested by ICEB to use a shear friction model ($\tau = m \tau_s$, $\tau$ being the shear friction stress, $\tau_s$ the material shear stress and m the friction factor) with m=1 [17]

Qform Extrusion uses friction model proposed by Levanov [18] on the contact part of workpiece surface:

$$f_\tau = m \cdot \bar{\sigma}/\sqrt{3}\,[1 - \exp(-1.25 \cdot \sigma_n/\bar{\sigma})] \quad (7)$$

where m is the friction factor and $\sigma_n$ is the normal contact pressure. Expression (7) can be considered as a combination of constant friction model and Coulomb friction model that inherits advantages of both ones. The second term in parenthesis takes into account the influence of normal contact pressure. For high value of contact pressure expression (7) provides approximately the same level of friction traction as constant friction model while for low contact pressure it gives friction traction that is approximately linearly dependent on normal contact stress.

**Table 5**
Dimensions and temperature of billet and tools

| | Diameter (mm) | Length (mm) | Temperature (ºC) |
|---|---|---|---|
| Ram | 209 | 1254 | 370 |
| Container | in=210 out=900 | 1055 | 420 |
| Billet | 203 | 500 | 480 |
| Die ring | out=530 in=282 | 154 | 450 |
| Mandrel | out=280 | 50 | 450 |
| Die | out=280 | 54 | 450 |
| Backer | out=280 | 50.5 | 450 |
| Bolster | out=530 | 250 | 430 |

Principal temperature parameters and boundary conditions of the process used in Qform Extrusion FEM simulation of die designs are shown in table 5. It shows principal dimensions of press elements and their temperature at the beginning of the extrusion process.

In addition, it should be noted that a velocity similar to those usually used for real extrusion tests on the press has been used as a speed condition for the process simulation [11]. The most common velocities in real tests of non-special profiles are usually between 120 and 170 mm/s at the exit of the press. For the simulations of this study, an output velocity value of 133 mm/s has been chosen.

Taking into account all these criterions and boundary conditions, two simulations of the extrusion have been carried out using Qform Extrusion software: the simulation of the initial design and the simulation of the design after the modifications.

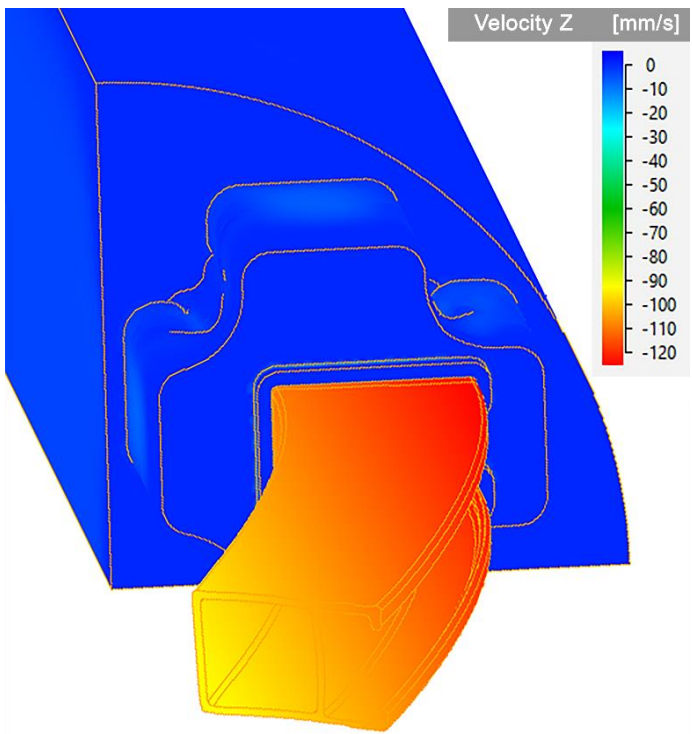

**Fig. 18.** Velocity Z QForm FEM extrusion simulation of initial design

Figure 18 shows the image of the velocity results in the extrusion direction (Z-direction) for the initial design. As can be seen, there is a clear imbalance in velocity in the Z-direction at the exit from the press.

Figure 19 shows the results of the velocity deviation from the average velocity for the initial design. The difference in velocity at the exit of the press is of the order of 18% between the slowest and fastest areas.

Such a large difference in velocity at the exit from the press causes, in the vast majority of cases, deformation of the profile and deviation of the mandrel. In addition, the deflections of the mandrel usually cause differences in thickness in the different hollow zones of the profile.

Experience indicates that it is very difficult to quantify what is the permissible velocity difference at the output of the press for each profile in order to obtain an extruded profile in accordance with established manufacturing tolerances. The admissible difference depends on many factors: rigidity of the profile itself, thickness of the profile, existence of special tolerances for the profile... However, it is generally accepted that minimizing speed differences always helps to ensure a result that is geometrically closer to the desired one.

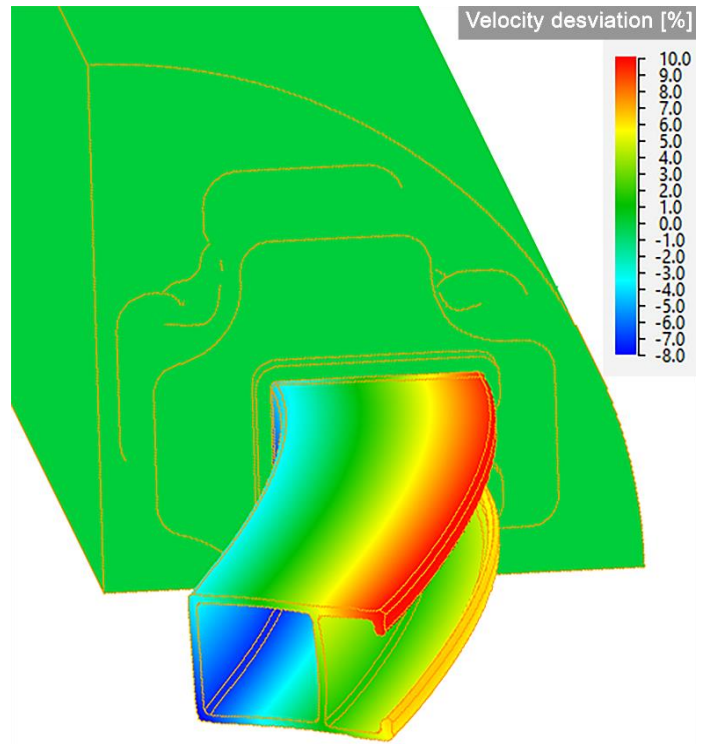

**Fig. 19.** Velocity deviation QForm FEM extrusion simulation of initial design

Figure 20 shows the image of the velocity results in the extrusion direction (Z-direction) for the final balanced design.

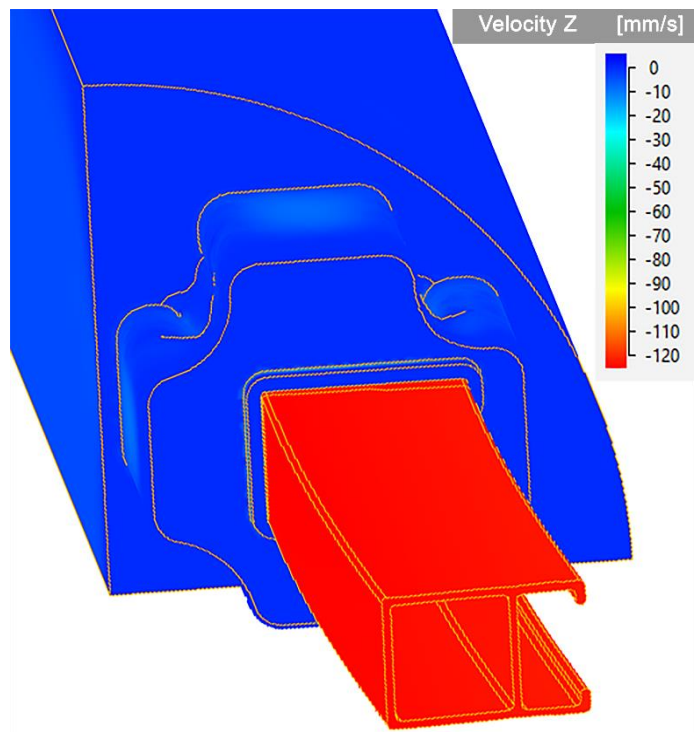

**Fig. 20.** Velocity Z QForm FEM extrusion simulation of final design

Figure 21 shows the results of the velocity deviation from the average velocity for the final design. The difference in velocity at the exit of the press is of the order of 2.2% between the slowest and fastest areas.

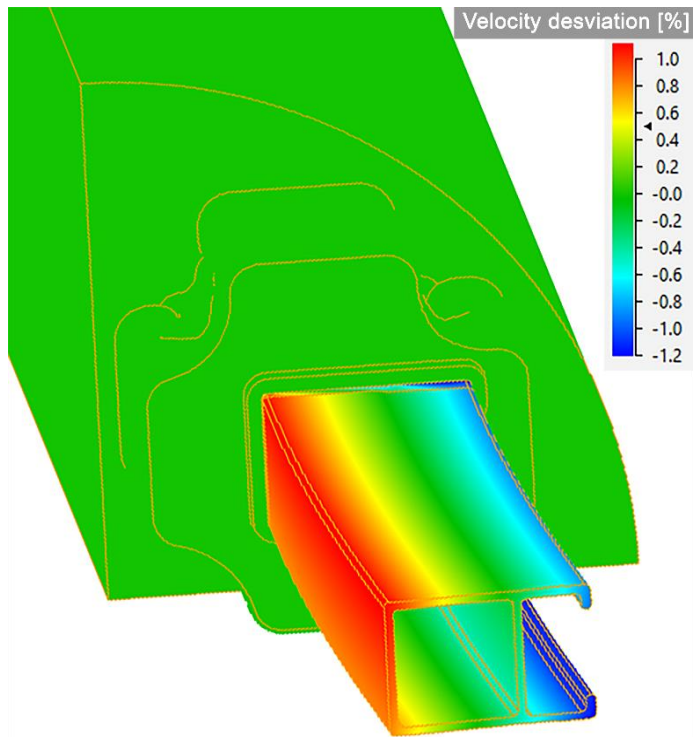

**Fig. 21.** Velocity deviation QForm FEM extrusion simulation of final design

In the dataset associated with this article [15] it is possible to see the nodal results and the process animations of the simulations process obtained by means of Qform Extrusion.

Another way to evaluate the goodness of the results obtained in the simulation of the extrusion is to use the root main square deviation (RMSC), or sum of squares of the nodal values of velocity deviation with respect to their mean value, from the bearing zone onwards. The aluminium zone before the bearing is not of interest for the analysis because the profile is not yet formed there.

**Table 6**
Root main square deviation (RMSC) for each design from bearing zone onwards

|  | Initial unbalanced design | Final balanced design |
|---|---|---|
| Mean nodal velocity deviation (m/s) | 0.00618 | 0.00115 |
| Sum of squares of nodal velocity deviation ($m^2/s^2$) | 274.57 | 5.53 |
| Root main square deviation – RMSC (m/s) | 16.57 | 2.35 |

Table 6 presents the values obtained in the root main square deviation for each of the designs. Both, the images of the results and the RMSC clearly show that the final design is much more balanced than the initial design.

## 5. Conclusions

The procedure followed to obtain the model for designing the ports of a specific type of porthole aluminium extrusion dies has, in general terms, achieved a very satisfactory result. The model developed can greatly facilitate the process of designing the ports for this type of die because it can serve as support for the designer at the beginning of the creation process. Until now, die design only depends on designer experience, the guidelines set by other similar dies with proven effectiveness and, in the best of cases, final adjustments based on extrusion simulation.

The model obtained is able to express through a mathematical expression the relationship between the main ports variables of a large number of proven effectiveness dies. Thus, it is a way of summarizing in a single expression the experience accumulated in a large number of designs over time.

At the same time the model obtained has limitations, one of them is that it is only valid for a very common type of dies: dies of four cavities with four ports each cavity. For other dies, with other numbers of cavities and ports it would be necessary in the future to obtain new models or to obtain a general model valid for all of them.

Therefore, an important line of research remains open in this sense because there is a great variety of possible port configurations and to achieve a similar verification tool would be a great advance compared to the current methodology.

Another limitation of the model is that it can only be used to check new designs because the geometric link between the different variables that make up the model makes it impossible for the model to serve directly for the creation of new designs. The model must be used in a loop design process, so that modifications are made to the design until the optimal design is reached. However, the model provides information on the direction and approximate magnitude order of the changes to be made to the design.

This use methodology opens another possible path for future development in the sense of automating the modification and iterative calculation of the model until obtaining a port design adjusted to the model. This automated procedure could be articulated around a parametric CAD (Computer Aided Design) tool that would modify the design incrementally until reaching an optimal result.

The effectiveness of the model could also be validated by manufacturing some of the dies whose designs have been based on this model. In this way, the check carried out by means of CAE simulation for aluminium extrusion would be complemented.

## Data availability

The raw/processed data required to reproduce statistical analysis and FEM simulation has been shared in Mendeley Data [15]: http://dx.doi.org/10.17632/ymrzd8j2r5.1

## CRediT authorship contribution statement

**Juan Llorca-Schenk**: Conceptualization, Methodology, Formal analysis, Investigation, Data curation, Writing - original

draft, Writing - review & editing, Project administration. **Irene Sentana-Gadea**: Validation, Writing - review & editing. **Miguel Sanchez-Lozano**: Writing - review & editing, Supervision.

**Declaration of competing interest**

The authors declare that they have no known competing financial interests or personal relationships that could have appeared to influence the work reported in this paper.

**Role of the funding source**

This research did not receive any specific grant from funding agencies in the public, commercial, or not-for-profit sectors.

**Acknowledgements**

This work was partially supported by the Design in Engineering and Technological Development Group (DIDET - Diseño en Ingeniería y Desarrollo Tecnológico) at the University of Alicante (UA VIGROB-032/19)

The authors thank the company Hydro Aluminium Extrusion Portugal HAEP, S.A. for facilitating and enabling access to feedback information on die performance during extrusion.

The authors would also like to thank QuantorForm Ltd for licensing Qform Extrusion for FEM verification